\documentclass[]{interact}

\usepackage{epstopdf}
\usepackage{subfigure}
\usepackage[caption=false]{subfig}
\usepackage{enumerate}
\usepackage{multirow}
\usepackage{multicol}
\usepackage[longnamesfirst,sort]{natbib}
\bibpunct[, ]{(}{)}{;}{a}{,}{,}
\renewcommand\bibfont{\fontsize{10}{12}\selectfont}

\theoremstyle{plain}

\theoremstyle{definition}

\theoremstyle{remark}

\usepackage{amsmath}
\usepackage{array}
\newcolumntype{P}[1]{>{\centering\arraybackslash}p{#1}}
\usepackage{graphicx}
\usepackage{algorithm}
\usepackage{algpseudocode}
\begin{document}

\articletype{Research Article}

\title{A Multi-objective Economic Statistical Design of CUSUM Chart: NSGA II Approach}

\author{
\name{Sandeep\textsuperscript{a}\thanks{CONTACT Sandeep. Email: sandeepkajal633@yahoo.com} and Arup Ranjan Mukhopadhyay\textsuperscript{a}}
\affil{\textsuperscript{a}Statistical Quality Control and Operations Research Unit, Indian Statistical Institute, 203, B.T.Road, Kolkata, 700108, West Bengal, India}
}

\maketitle

\begin{abstract}
This paper presents an approach for the economic statistical design of the Cumulative Sum (CUSUM) control chart in a multi-objective optimization framework. The proposed methodology integrates economic considerations with statistical aspects to optimize the design parameters like the sample size ($n$), sampling interval ($h$), and decision interval ($H$) of the CUSUM chart. The Non-dominated Sorting Genetic Algorithm II (NSGA II) is employed to solve the multi-objective optimization problem, aiming to minimize both the average cost per cycle ($C_E$) and the out-of-control Average Run Length ($ARL_\delta$) simultaneously. The effectiveness of the proposed approach is demonstrated through a numerical example by determining the optimized CUSUM chart parameters using NSGA-II. Additionally, sensitivity analysis is conducted to assess the impact of variations in input parameters. The corresponding results indicate that the proposed methodology significantly reduces the expected cost per cycle by about 43\% when compared to the findings of the article by M. Lee in year 2011. No comparison of $ARL_\delta$ could be made due to the non-availability of references. This highlights the practical relevance and potential of this study for the right application of the technique of the CUSUM chart for process control purposes in industries.
\end{abstract}

\section{Introduction}
Control charts are known to be a pivotal tool for exercising control over a process in statistical process control, with their origins dating back to Shewhart's pioneering work in the 1920s. The design of the control chart encompasses the critical task of selecting parameters such as sample size, sampling interval, and control limits' multiplier. These design parameters are crucial as they directly influence the quality of the control over the underlying process being monitored.

The cumulative sum (CUSUM) control chart, pioneered by Page in 1954, is a widely adopted method for exercising control over the mean of a quality characteristic in production processes. The usefulness of the techniques of the CUSUSM control chart in practical applications in industry is to identify the trend of the process. If the trend is found to worsen, the concerned quality characteristic needs to be analyzed for possible assignable or special causes with or without halting the process. One can refer to the article written by \cite{2001} where the CUSUM chart has been used for routine bias correction on the shop floor for measuring and controlling moisture content in tobacco. When compared with the $\bar{X}$-chart, the CUSUM chart demonstrates superior efficiency in detecting small to moderate shifts in the process mean. Similar to the Shewhart control chart the CUSUM control chart also needs to be implemented by collecting samples at regular intervals. The chart depicts a cumulative sum of deviations between sample means and a target value over time. As long as the computed CUSUM statistic remains within a predefined decision interval, the process is said to be in control. However, if the CUSUM statistic exceeds the decision interval, it serves as an indication that the process went out of control. Consequently, it requires further investigation to find the potential causes (assignable causes) of this change. However, there are some costs associated with finding the assignable causes, like the costs of sampling, the costs of eliminating the assignable causes, and the costs of producing nonconforming units. It is interesting to note that the cost of producing nonconforming units increases with less sampling and less effort or lower costs of eliminating assignable causes. This underscores the importance of the trade-off between sampling cost followed by the cost of eliminating assignable causes and the cost of producing nonconforming units. From an economic viewpoint, it is always better to consider the economic statistical design of the control chart. 

The economic design of control charts has always been a significant area to focus on in statistical process control. The objective of economic design is to determine optimal design parameters that minimize the expected cost per hour, thereby achieving control over a process in a cost-effective manner. \cite{1956} introduced the first cost model for $\bar{X}$ chart, paving the way for subsequent research in the economic design of other control charts. A few of these are mentioned here. \cite{1986} proposed the unified economic design of control charts. The economic design of the control chart for sustainable operations under the gamma shock model was proposed by \cite{2020}. \cite{Y2021} proposed a novel run rules-based MEWMA scheme for monitoring general linear profiles. Detection of intermittent faults based on moving average $T^2$ control charts with multiple window lengths was studied by \cite{Z2020}. The optimal design of $S^2$-EWMA control chart for short production runs is given by \cite{2023}. \cite{Z2024} proposed the joint design of VSI $T_r$ control chart. \cite{2024} designed a log-logistic-based EWMA control chart using MOPSO and VIKOR approaches for monitoring cardiac surgery performance. \cite{H2024} proposed combined Shewhart-EWMA and Shewhart-CUSUM monitoring schemes for time between events.

Four parameters are of paramount importance in designing a CUSUM chart. These parameters are the optimal sample size ($n$), the optimal sampling interval ($h$), the reference value ($K$), and the decision interval ($H$). \cite{1968} was the first researcher to explore the economic design of the CUSUM chart, laying the foundation for subsequent investigations in this area. Since then various methodologies have been proposed for the economic design of the CUSUM chart. For instance, \cite{1973} developed a procedure specifically tailored for controlling the mean of a process that follows a normal distribution using a CUSUM chart.

\cite{1974} introduced a production model centered on quality surveillance, employing the CUSUM chart with decision interval criteria. \cite{1992} devised a search algorithm utilizing the one-dimensional $H$-pattern search technique given by \cite{1961} for the economic design of the CUSUM chart. \cite{1995} utilized two-level fractional factorial designs to find the optimal parameters using the economic model for control charts given by \cite{1986} under CUSUM conditions.

\cite{2005} proposed an innovative approach for monitoring and evaluating environmental performance through the economic design of the CUSUM chart. Additionally, an economic model of the CUSUM chart for controlling the process mean in short production runs was proposed by \cite{2006}. \cite{2011} proposed the economic design of a CUSUM control chart for non-normally correlated data. Overall, the economic design of the CUSUM chart has drawn significant attention and continues to be a subject of active research, with various methodologies and applications contributing to its advancement in the field of statistical process control.

Along with the economic aspects associated with control charts, sometimes it is necessary to consider the statistical aspects like detecting the shift as early as possible. The multi-objective economic statistical design of control charts is one way to study the economic aspects along with the statistical aspects. In the literature, there are several papers dealing with the multi-objective economic design of other control charts, some of which are included here. \cite{1999} studied the multi-objective economic design of an $\bar{X}$ control chart. \cite{Y2012} used a multi-objective particle swarm optimization algorithm to develop a multi-objective model for the optimal design of $\bar{X}$ and S control charts. \cite{2012} studied the multi-objective economic statistical design of $\bar{X}$ control chart considering Taguchi loss function. \cite{2013} examined a bi-objective optimization model for the economic-statistical design of control charts. \cite{2014} developed a multi-objective optimization model for the optimal design of a c-chart. \cite{B2014} proposed a multi-objective economic-statistical design for the cumulative count of conforming control chart. \cite{Z2015} presented a multi-objective optimization model for designing a control chart with fuzzy parameters to monitor the process mean.

However, the multi-objective economic statistical design of the CUSUM chart is yet to be explored. So, in this article, a multi-objective economic statistical design of the CUSUM chart is proposed using the cost model given by \cite{1986}. In this model, we would try to minimize the expected cost per cycle ($C_E$) as well as the out-of-control average run length ($ARL_\delta$) while maintaining a reasonably large in-control average run length ($ARL_0$). Along with minimizing the two objectives, we would find out the optimal parameters $n$, $h$, and $H$. The optimal value of $K$ is the half of the magnitude of the shift given in $\sigma$ units, as mentioned in Section \ref{Section 2}. The proposed multi-objective model has been solved with the help of Non-dominated Sorting Genetic Algorithm II (NSGA II) which was introduced by \cite{K2011}. NSGA-II has been chosen for this study because it was designed specifically to handle multi-objective problems. It uses a genetic algorithm framework to seek out Pareto-optimal solutions, which are characterized by the inability to enhance one objective without compromising another. This enables decision-makers to navigate a spectrum of alternatives while weighing the trade-offs between different objectives. By harnessing evolutionary search techniques like crossover and mutation, NSGA II efficiently traverses the solution space, making it adept at handling intricate optimization problems with numerous decision variables and constraints, mirroring those found in our proposed model. Furthermore, NSGA II possesses adaptability, allowing it to adjust to changes in problem formulations or objective functions with minimal alterations to the algorithm. This attribute lends itself to flexibility, enabling the seamless incorporation of additional objectives or constraints as the problem evolves over time. \cite{A2014} considered a multi-objective economical-statistical design of the EWMA chart and solved it using NSGA II and MOGA algorithms. \cite{M2015} used NSGA-II algorithm to solve multi-objective design of $\bar{X}$ control chart. \cite{2019} proposed the economic-statistical design of the c-chart with multiple assignable causes and solved it using a hybrid NSGA-II approach.

We are using NSGA II since it is specifically designed to handle problems with multiple objectives. It can handle complex optimization problems with many decision variables and constraints as we have in our proposed model. It can adapt to changes in problem formulations or objective functions without significantly modifying the algorithm. It offers flexibility in incorporating additional objectives or constraints as the problem evolves. In contrast, the primary aim of goal programming is minimizing deviations from predefined goals and may not be as effective in handling multiple conflicting objectives. Goal programming typically produces a single solution that minimizes deviations from goals but may not provide a comprehensive view of the problem's trade-offs. It may face challenges in dealing with intricate constraints. It may require adjustments in the formulation or constraints when objectives or priorities change. Moreover, solving a double-objective problem by using goal programming requires users to provide a weight factor for each goal. The resulting solution, therefore, depends on the chosen set of weight factors. Also, the goal programming has difficulty in finding solutions for the problems having non-convex feasible decision space as shown in \cite{D1999}.

The rest of the paper is organized like this. The sample statistic of the CUSUM chart is given in Section \ref{Section 2}. The formulation of the multi-objective economic statistical design of the CUSUM chart is given in Section \ref{Section 3}. The pseudocode for using the NSGA II is given in Section \ref{Section 4}. The aspects of applicability for the proposed approach are demonstrated by a numerical example in Section \ref{Section 5}. The results of the sensitivity analysis are provided in Section \ref{Section 6}.  Benefits and practical applications are discussed in Section \ref{Section 7}. The conclusion for the proposed approach and the scope for future work are given in Section \ref{Section 8}.

\section{The Statistic of CUSUM chart}\label{Section 2}
Let's consider a scenario where the variation of a quality characteristic $x$ follows the normal distribution with mean $\mu$ and standard deviation $\sigma$ in the in-control state, denoted as $x \sim N(\mu,\sigma^2)$, where both $\mu$ and $\sigma$ are known. However, over time, the process may transit to an out-of-control state, resulting in a shift in the mean of the quality characteristic from $\mu_0$ (where $\mu_0$ = $\mu$) to $\mu_0\pm\delta\sigma_0$, where $\mu_0$ and $\sigma_0$ represent respectively the sample mean and the sample standard deviation. Here, $\delta$ signifies the magnitude of the shift in mean, while the standard deviation is assumed to remain constant. Samples of size $n$ are taken after every $h$ hours of production, with reference value ($K$) and decision interval ($H$). Subsequently, the gathered sample information is plotted on a CUSUM chart. In the event of CUSUM statistic surpassing the decision interval, an investigation to identify and eliminate the assignable cause is initiated. The CUSUM statistic can be calculated as
\begin{equation}\label{equ1}
\begin{aligned}
    C_i^- = max(0, (\mu_o - K) - x_i + C_{i-1}^-)\\
    C_i^+ = max(0, x_i - (\mu_o + K) + C_{i-1}^+)    
\end{aligned}
\end{equation}
where i denote the $i^{th}$ sample and $ C_0^-$ = $ C_0^+$ = 0. 

The effectiveness of a control chart can be assessed using the average run length ($ARL$), which represents the average number of samples needed to detect an out-of-control condition or trigger a false alarm. The in-control $ARL$ ($ARL_0$) is used for calculating the false alarm rate whereas the out-of-control $ARL$ ($ARL_\delta$) is an indicator of the power (or effectiveness) of the control chart. One of the major difficulties in the economic design of the CUSUM chart is the evaluation of average run lengths. In this paper, we are utilizing approximation for calculating $ARL$ given by \cite{1985}. The main reason for using Siegmund's approximation is its simplicity. \cite{1993} also recommended to use approximation given by \cite{1985}. For one-sided CUSUM (i.e., $C_i^-$ or $C_i^+$) Siegmund's approximation for $ARL$ is
\begin{equation}\label{equ2}
    ARL = \frac{e^{-2\Delta b} + 2\Delta b - 1}{2\Delta^2}
\end{equation}
For $\Delta\neq 0$, where $\Delta = -\delta - K$ for the lower one-sided CUSUM $C_i^-$, $\Delta = \delta - K$ for the upper one-sided CUSUM $C_i^+$, $\delta$ being the magnitude of process shift in $\sigma$ units for which $ARL$ needs to be calculated, $K = \frac{\delta}{2}$, and $b = H + 1.166$. For $\Delta = 0$, $ARL$ can be calculated by $b^2$. Hence, the formula given in equation (\ref{equ2}) can be used to determine $ARL_0$ when $\delta = 0$ and when $\delta\neq 0$ it can be used to calculate $ARL_\delta$.

For $\delta = 0$, equation (\ref{equ2}) can be used for calculating one-sided in-control $ARL$ (i.e., $ARL_0^-$ and $ARL_0^+$):
\begin{equation}\label{equ3}
    ARL_0^- = ARL_0^+ = \frac{e^{2Kb} - 2Kb - 1}{2K^2}
\end{equation}

Whereas for $\delta\neq 0$, equation (\ref{equ2}) can be used for calculating one-sided out-of-control $ARL$ (i.e., $ARL_\delta^-$, and $ARL_\delta^+$)
\begin{equation}\label{equ4}
    \begin{aligned}
    ARL_\delta^-=\frac{e^{2\delta b + 2Kb} - 2\delta b - 2Kb - 1}{2(\delta + K)^2}\\
    ARL_\delta^+=\frac{e^{-2\delta b + 2Kb} + 2\delta b - 2Kb - 1}{2(\delta - K)^2}
    \end{aligned}
\end{equation}

For calculating $ARL$ of a two-sided CUSUM, one can use the formula given in equation (\ref{equ5}) using two one-sided $ARLs$ (i.e., $ARL^-$, and $ARL^+$)

\begin{equation}\label{equ5}
    \frac{1}{ARL} = \frac{1}{ARL^-} + \frac{1}{ARL^+}
\end{equation}

The above values of $ARL_0$ and $ARL_\delta$ are used in the multi-objective economic statistical design of the CUSUM chart defined in Section \ref{Section 3}.

\section{A Multi-objective Economic Statistical Design of CUSUM chart}\label{Section 3}
In this section, based on certain assumptions the expected cost per cycle has been considered for proposing a multi-objective economic statistical design of the CUSUM chart.

\subsection{Assumptions for the Model}

The following assumptions are deemed valid for the proposed model:

(1) The mean of the quality characteristics is assumed to follow Normal distribution.

(2) The mean of the quality characteristic shifts from $\mu_0$ to $\mu_0 + \delta\sigma_0$

(3) The occurrence of an assignable cause follows an exponential distribution with a mean of $1/\lambda$. 

\subsection{Expected Cost Per Cycle}
In this article, \cite{1986} cost model has been extended to a multi-objective economic statistical design of a CUSUM chart. The reason is that it is the most widely used statistically constrained economic model. The expected cost per cycle($C_E$) is:
\begin{eqnarray}\label{equ6}
    C_E &=& \frac{\frac{C_0}{\lambda} + C_1(-\tau + nt + h(ARL_\delta) + \gamma_1T_1 + \gamma_2T_2) + \frac{SW}{ARL_0} + Y}{\frac{1}{\lambda}+\frac{(1-\gamma_1)ST_0}{ARL_0}-\tau + nt + h(ARL_\delta) + T_1 + T_2}\nonumber\\
    &&+\frac{(\frac{d+n y}{h})(\frac{1}{\lambda}-\tau + nt + h(ARL_\delta) + \gamma_1T_1 + \gamma_2T_2)}{\frac{1}{\lambda}+\frac{(1-\gamma_1)ST_0}{ARL_0}-\tau + nt + h(ARL_\delta) + T_1 + T_2}
\end{eqnarray}
The parameters given in equation (\ref{equ6}) are defined below:\\
$C_0$: Quality cost per hour for the in-control process\\
$C_1$: Quality cost per hour for the out-of-control process\\
$\tau$: Average time taken for assignable cause to occur and can be determined by $\tau$:
\begin{eqnarray}\label{equ7}
    \tau &=&\frac{\int_{rh}^{(r+1)h}e^{-\lambda t}\lambda (t-rh)dt}{\int_{rh}^{(r+1)h}e^{-\lambda t}\lambda dt}\nonumber\\
        &=& \frac{1-(1+\lambda h)e^{-\lambda h}}{\lambda(1-e^{-\lambda h})}\nonumber\\
        &=& \frac{1}{\lambda} - \frac{h}{e^{\lambda h}-1}
\end{eqnarray}
$t$: Average time to take a sample and obtain the results\\
$ARL_0$: Average in-control run length\\
$ARL_\delta$: Average out-of-control run length\\
$T_0$: Average time associated with a false alarm\\
$T_1$: The average time required to discover an assignable cause\\
$T_2$: The average time required to eliminate an assignable cause\\
$\gamma_1$: A binary variable that takes the value 1 if the production continues during the search for an assignable cause and 0 otherwise\\
$\gamma_2$: A binary variable that takes the value 1 if the production continues during the elimination of an assignable cause through intervening in the process and 0 otherwise\\
$S$: Average number of samples taken while the process is in control. It can be determined by:
\begin{eqnarray}\label{equ8}
S & = &\sum_{r=0}^\infty\int_{rh}^{(r+1)h}r\lambda e^{-\lambda t}dt \nonumber\\
& = & \frac{1}{e^{\lambda h} - 1}
\end{eqnarray}
$W$: The average cost for searching an assignable cause when there is none\\
$Y$: The average cost of identifying and eliminating an assignable cause\\
$d$: The cost per sample for maintaining the CUSUM chart in a process\\
$y$: The variable cost of sampling an inspection unit\\
\subsection{Multi-objective Economic Statistical Design}
A multi-objective design of a CUSUM chart with two objectives has been proposed in this article. Our aim is to minimize the expected cost per cycle ($C_E$) as well as the Average out-of-control Run Length ($ARL_\delta$). The multi-objective design is given as follows:\\
\begin{eqnarray}\label{equ9}
    Min\,\, C_E\nonumber\\
    Min\,\, ARL_\delta\nonumber\\
    s.t.\,\,\,\,\,\,ARL_0 \geq ARL_L\\
    ARL_\delta \leq ARL_u\nonumber\\
    and \,\, n \in Z^+\nonumber
\end{eqnarray}
$ARL_L$ and $ARl_u$ are the respective lower and upper bounds of $ARL_0$ and $ARL_\delta$.

It has already been mentioned that to arrive at the economic statistical design of a CUSUM chart, one requires three decision variables, namely, $n$, $h$, and $H$. This article uses the NSGA II method to determine the optimal parameters of the multi-objective design for the CUSUM chart given in equation (\ref{equ9}).

\section{Pseudo Code for Using NSGA II}\label{Section 4}
At iteration \( k \), NSGA II initializes with a population \( P_k \) consisting of \( N \) candidate solutions. It then proceeds to a loop where \( N \) children are generated. Each child is created by selecting a pair of parents through binary tournament selection, with the criterion being the Crowded-comparison Operator (CCO) operator. The CCO compares two solutions based on their Pareto rank and crowding distance, which are described below.

\textbf{Pareto Rank}: A solution with a lower rank (i.e., it belongs to a better Pareto front) is considered superior.

\textbf{Crowding Distance}: If two solutions have the same rank, the one with the higher crowding distance (i.e., it is more isolated from others) is preferred to maintain diversity.

The parents chosen using the CCO criterion then undergo crossover to produce offspring, which subsequently undergoes mutation. All children are stored in a matrix \( Q_k \).

A combined population \( R_k = P_k \cup Q_k \) is formed. Utilizing fast non-dominated sorting, the Pareto fronts \( F_0, F_1, \ldots \) of \( R_k \) are determined. Subsequently, the population for the next generation, \( P_{k+1} \), is constructed.

\( P_{k+1} \) is initially an empty set. The algorithm then sets \( P_{k+1} = P_{k+1} \cup F_0 \). If the cardinality of \( P_{k+1} \), denoted as \( |P_{k+1}| \), equals \( N \), the creation of \( P_{k+1} \) is concluded, and the algorithm proceeds to the next iteration. However, if \( |P_{k+1}| < N \), the creation process continues. In this case, fronts are added to \( P_{k+1} \) in the order of their ranking until \( |P_{k+1}| = N \).

While incorporating fronts into \( P_{k+1} \), it's possible that one front, denoted as \( a_{\text{last}} \), may not entirely fit into \( P_{k+1} \). In such instances, the solutions of \( F_{\text{last}} \) are arranged in descending order of the CCO, and these ordered solutions are sequentially inserted into \( P_{k+1} \) until \( |P_{k+1}| = N \). The decision to conclude or continue forming \( P_{k+1} \) based on its cardinality relative to \( N \) is essential for maintaining the appropriate working of the algorithm for consistent and valid results ensuring that the population remains fixed in size while also being of high quality and diversity.

The pseudocode for implementing NSGA-II is given in Algorithm \ref{alg:NSGA-II-full}. This algorithm was given by \cite{S2021}. Following this algorithm, the code to implement NSGA II has been appropriately modified for our proposed model. In order to write the code for implementing NSGA II in the realm of the CUSUM chart, certain pseudocodes are required. These pseudocodes pertaining to Pareto Ranking, Fast Non-dominated Sorting, and Crowded-comparison Operator are also given in \cite{S2021}.
\begin{algorithm}[ht!]\scriptsize{}
\caption{Non-dominated Sorting Genetic Algorithm II (NSGA-II)}
\label{alg:NSGA-II-full}
\begin{algorithmic}[1]
    \State $k \leftarrow 0$ \Comment{index of the current iteration}
    \State $P_k \leftarrow$ initial population of N candidate solutions \Comment{satisfying constraints}
    \State $Q_k \leftarrow \phi$ \Comment{population of children}
    \While {termination criterion not satisfied}
        \For {$i \in \{1,...,N-1\}$}
            \State $(\mathbf{p_1},\mathbf{p_2}) \leftarrow$ select two parents using binary tournament from $P_k$
            \State $\mathbf{r} \leftarrow$ recombine($\mathbf{p_1},\mathbf{p_2})$ \Comment{create a child by crossover}
            \State $\mathbf{q} \leftarrow$  mutate($\mathbf{r}$) \Comment{mutate the child}
            \State $Q_k \leftarrow$ $Q_k \cup \{\mathbf{q}\}$ \Comment{update the population of children}
        \EndFor
        \State $R_k \leftarrow P_k \cup Q_k$ \Comment{create a combined population}
        \State $\mathcal{F} \leftarrow$ all non-dominated fronts in $R_k$ \Comment{$\mathcal{F} = \{\mathcal{F}_0,\mathcal{F}_1,...\}$}
        \State $P_{k+1} \leftarrow \{\}$ \Comment{new population}
        \While{$P_{k+1}$ does not have N individuals}
        \State $i \leftarrow 0$
        \If {$P_{k+1}$ has room for all elements of $\mathcal{F}_i$}
        \State $P_{k+1} \leftarrow P_{k+1} \cup \mathcal{F}_i$ \Comment{add i-th front to the parent population}
        \Else
        \State $P_{k+1} \leftarrow P_{k+1} \cup$ the first $(N - |P_{k+1}|)$ points of $\mathcal{F}_i$
        \EndIf
         \State $i \leftarrow i+1$
        \EndWhile
        \State delete the fronts which could not be inserted in $P_{k+1}$
        \State $k \leftarrow k+1$
    \EndWhile 
\end{algorithmic}
\end{algorithm}

It is worthwhile to mention here that NSGA II generates a set of optimal solutions that are represented as Pareto fronts. These fronts are a key concept in multi-objective optimization, particularly in evolutionary algorithms like NSGA-II. These fronts are a visual and conceptual representation of the trade-offs between multiple conflicting objectives. Each front contains solutions that are optimal in the sense that no other solution is strictly better in all objectives. The shape and distribution of solutions on the Pareto fronts provide valuable insights for decision-making in multi-objective optimization problems.

\section{Numerical Example}\label{Section 5}
In this section, an application of the economic design of the CUSUM chart is shown with the help of an example taken from \cite{2011} where he considered a hypothetical set of process and cost parameters.

\begin{figure}[h!]
    \centering
    \includegraphics[width=12.4cm,height=7cm]{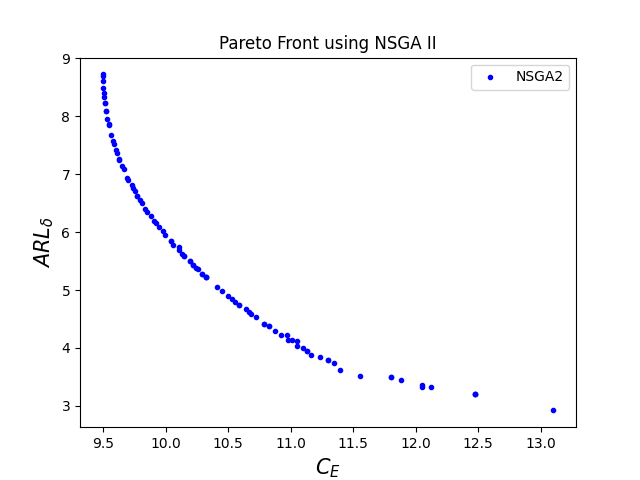}
    \caption{Optimal Pareto Front}
    \label{figure 1}
\end{figure}

He considered a factory in which yogurt drinks are produced and contained in bottles. The target quantity of yogurt drink for each bottle is 0.02 liters. The produced yogurt drink is then inserted into fifteen bottles at a time. Now, those fifteen bottles are packed in a box later. Suppose that the hourly in-control quality cost is $C_0$ = \$10 and that of an out-of-control state is $C_1$ = \$100. Since the inter-occurrence time of the assignable causes was assumed to follow an exponential distribution, let's assume that assignable causes occur with a frequency of about one every hundred hours of operation. Thus, $\lambda$ = 0.01. The cost per sample for maintaining the CUSUM chart and the variable cost of sampling respectively are $d$ = \$0.5 and $y$ = \$0.1. The cost of investigating a false alarm is $W$ = \$50. The average cost of identifying and eliminating an assignable cause is $Y$ = \$25. It takes on an average of three minutes ($t$ = 0.05 hours) to take a sample and obtain the results. It requires about $T_1$ = 2 hours to discover an assignable cause, and it requires about $T_2$ = 2 hours to eliminate the assignable cause. The lower and upper control limits on $ARL_0$ and $ARL_\delta$ respectively are 200 and 14 units and constraints on $n$, $h$, and $H$ respectively range from 2 to 20, 0.01 to 2, and 0.0001 to 5. Further, it is assumed that the process continues to operate while searching and elimination of an assignable cause are going on. There are 82 non-dominated solutions, so it will be difficult to give all solutions in the form of a table. Due to this reason, table \ref{table 1} contains the solution in terms of percentile (i.e. 5th, 10th, 15th,....,100th) with an increment of 5 percentile beginning with the 1st percentile. The corresponding optimal Pareto Front is shown in figure \ref{figure 1}.

\begin{table}[ht]
    \centering
    \caption{The non-dominated set of optimal parameters along with $C_E$ and $ARL_\delta$.}
     \begin{tabular}{|P{2.4cm}|P{2.4cm}|P{2.4cm}|P{2.4cm}|P{2.5cm}|}\hline
  $C_E$ & $ARL_\delta$ & $n$ & $h$ & $H$\\\hline
 9.50 & 8.72 & 2 & 0.36 & 4.19\\
 9.50 & 8.49 & 2 & 0.37 & 4.07\\
 9.52 & 8.10 & 2 & 0.40 & 3.88\\
 9.56 & 7.69 & 2 & 0.43 & 3.67\\
 9.61 & 7.37 & 2 & 0.44 & 3.51\\
 9.69 & 6.94 & 2 & 0.48 & 3.29\\
 9.75 & 6.71 & 2 & 0.53 & 3.18\\
 9.75 & 6.71 & 2 & 0.53 & 3.18\\
 9.84 & 6.39 & 2 & 0.55 & 3.02\\
 9.92 & 6.15 & 2 & 0.56 & 2.89\\
 10.04 & 5.84 & 2 & 0.65 & 2.74\\
 10.13 & 5.63 & 2 & 0.64 & 2.63\\
 10.24 & 5.38 & 2 & 0.71 & 2.50\\
 10.41 & 5.05 & 2 & 0.78 & 2.33\\
 10.55 & 4.80 & 2 & 0.82 & 2.20\\
 10.69 & 4.58 & 2 & 0.94 & 2.09\\
 10.92 & 4.23 & 2 & 1.07 & 1.90\\
 11.05 & 4.13 & 2 & 0.94 & 1.85\\
 11.16 & 3.88 & 2 & 1.15 & 1.73\\
 11.40 & 3.62 & 2 & 1.19 & 1.59\\
 12.05 & 3.37 & 2 & 1.03 & 1.46\\
 13.10 & 2.92 & 2 & 1.07 & 1.22\\\hline
    \end{tabular}
    \label{table 1}
\end{table}

The minimum value of expected cost per cycle ($C_E$) is \$9.50 and the pertinent optimal parameters are $n$ = 2, $h$ = 0.36 hours, and $H$ = 4.19 units. \cite{2011} obtained the corresponding optimal values of $C_E$, $n$, $h$, and $H$ under the normality condition respectively as \$16.78, 2, 0.85 hours, and 1.69 units where he used Markov chain-based approach given by \cite{1997} to find $ARL_0$ and $ARL_\delta$. If one uses the proposed multi-objective economic statistical approach, the expected cost per cycle ($C_E)$ decreases by 43.39\%.

\section{Sensitivity Analysis}\label{Section 6}
Table \ref{table 2} contains two sets of values along with optimal parameters $n$, $h$, and $H$ for a given $\delta$ and these are (i) $C_E$ is minimum and $ ARL_\delta$ is maximum (ii) $C_E$ is maximum and $ ARL_\delta$ is minimum. As the magnitude of the shift ($\delta$) increases, both $C_E$ and $ARL_\delta$ decrease. However, $n$ remains the same. As the magnitude of shift increases, $h$ decreases, and $H$ increases.
\begin{table}[ht!]
    \centering
    \caption{Optimal values of $C_E$ and $ARL_\delta$ along with respective optimal parameters for different values of $\delta$.}
    \begin{tabular}{|P{2.5cm}|P{2.5cm}|P{2.5cm}|P{1.6cm}|P{1.45cm}|P{1.45cm}|}\hline
        $\delta$ & $C_E$ & $ARL_\delta$ & $n$ & $h$ & $H$\\\hline
        \multirow{2}{*}{$1.0$} & 9.50 & 8.72 & 2 & 0.36 & 4.19\\
        & 13.10 & 2.92 & 2 & 1.07 & 1.22\\\hline
        \multirow{2}{*}{$1.5$} & 7.97 & 4.78 & 2 & 0.48 & 3.08\\
        & 10.91 & 2.03 & 2 & 1.09 & 1.00\\\hline
        \multirow{2}{*}{$2.0$} & 7.09 & 2.96 & 2 & 0.64 & 2.30\\
        & 9.34 & 1.48 & 2 & 1.19 & 0.80\\\hline
        \multirow{2}{*}{$2.5$} & 6.50 & 2.05 & 2 & 0.79 & 1.80\\
        & 8.34 & 1.08 & 2 & 1.46 & 0.57\\\hline
    \end{tabular}
    \label{table 2}
\end{table}

Table \ref{table 3} also contains two sets of values along with optimal parameters $n$, $h$, and $H$ when cost parameters are increased from their respective low levels to high levels with a change of one-factor-at-a-time and these are (i) $C_E$ is minimum and $ ARL_\delta$ is maximum (ii) $C_E$ is maximum and $ ARL_\delta$ is minimum. All the low levels of input parameters are kept the same as given in the numerical example section and high levels are chosen at random since it is a common practice in research articles that deal with the economic design of control charts. For  $\delta$ = 1.0, table \ref{table 2} presents the corresponding values of the optimal parameters when each input variable is at its pertinent low level. The minimum change in the minimum $C_E$ is obtained corresponding to the parameter $C_0$ and the maximum change in the minimum $C_E$ is obtained corresponding to the parameter $C_1$. Similarly, The minimum change in the maximum $C_E$ is also obtained corresponding to the parameter $C_0$, and the maximum change in the maximum $C_E$ is obtained corresponding to the parameter $C_1$. Also, the minimum change in the minimum $ARL_\delta$ is obtained corresponding to the parameter $C_0$, and the maximum change in the minimum $ARL_\delta$ is obtained corresponding to the parameter $d$. Similarly, The minimum change in the maximum $ARL_\delta$ is also obtained corresponding to the parameter $C_0$, and the maximum change in the maximum $ARL_\delta$ is obtained corresponding to the parameter $d$.
\begin{table}[ht!]
    \centering
    \caption{Optimal values of $C_E$ and $ARL_\delta$ along with respective optimal parameters for cost parameters.}
    \begin{tabular}{|P{2cm}|P{1.4cm}|P{1.4cm}|P{1.25cm}|P{1.25cm}|P{1.25cm}|P{1.25cm}|P{1.25cm}|}\hline
        Input Parameters & Low Level & High Level & $C_E$ & $ARL_\delta$ & $n$ & $h$ & $H$\\\hline
        \multirow{2}{*}{$C_0$} & 10 & 20 & 9.59 & 8.83 & 2 & 0.36 & 4.25\\
        & & & 13.02 & 3.12 & 2 & 0.96 & 1.32\\\hline
        \multirow{2}{*}{$C_1$} & 100 & 200 & 15.64 & 8.92 & 2 & 0.24 & 4.29\\
        & & & 20.96 & 3.42 & 2 & 0.52 & 1.48\\\hline
        \multirow{2}{*}{$W$} & 50 & 100 & 9.91 & 9.92 & 2 & 0.34 & 4.79\\
        & & & 20.77 & 3.18 & 2 & 0.77 & 1.36\\\hline
        \multirow{2}{*}{$Y$} & 25 & 50 & 9.73 & 8.52 & 2 & 0.37 & 4.09\\
        & & & 13.80 & 3.05 & 2 & 0.89 & 1.29\\\hline
        \multirow{2}{*}{$d$} & 0.5 & 5 & 14.15 & 3.57 & 2 & 1.87 & 1.56\\
        & & & 19.68 & 1.69 & 2 & 1.60 & 0.54\\\hline
        \multirow{2}{*}{$y$} & 0.1 & 1 & 12.38 & 5.17 & 2 & 1.02 & 2.39\\
        & & & 15.74 & 2.17 & 2 & 1.45 & 0.81\\\hline
    \end{tabular}
    \label{table 3}
\end{table}

Table \ref{table 4} contains two sets of values along with optimal parameters $n$, $h$, and $H$ when time-related parameters are increased from their respective low levels to high levels with one-factor-at-a-time change. The minimum change in the minimum $C_E$ is obtained corresponding to the parameter $t$ and the maximum change in the minimum $C_E$ is obtained corresponding to the parameter $\lambda$. Similarly, the minimum change in the maximum $C_E$ is also obtained corresponding to the parameter $t$, and the maximum change in the maximum $C_E$ is obtained corresponding to the parameter $\lambda$. Also, the minimum change in the minimum $ARL_\delta$ is obtained corresponding to the parameters $T_1$ and $T_2$, and the maximum change in the minimum $ARL_\delta$ is obtained corresponding to the parameter $\lambda$. Similarly, the minimum change in the maximum $ARL_\delta$ is obtained corresponding to the parameter $t$, and the maximum change in the maximum $ARL_\delta$ is obtained corresponding to the parameter $\lambda$. The optimum values remain the same when $T_0$ is changed from low to high level since the process is assumed to operate continuously while the search and elimination of an assignable cause are going on. All of the corresponding optimal Pareto fronts are given in the Appendix.
\begin{table}[ht!]
    \centering
    \caption{Optimal values of $C_E$ and $ARL_\delta$ along with respective optimal parameters for time parameters.}
    \begin{tabular}{|P{2cm}|P{1.4cm}|P{1.4cm}|P{1.25cm}|P{1.25cm}|P{1.25cm}|P{1.25cm}|P{1.25cm}|}\hline
        Input Parameters & Low Level & High Level & $C_E$ & $ARL_\delta$ & $n$ & $h$ & $H$\\\hline
        \multirow{2}{*}{$\lambda$} & 0.01 & 0.05 & 28.10 & 8.23 & 2 & 0.20 & 3.94\\
        & & & 34.45 & 2.40 & 2 & 0.71 & 0.94\\\hline
        \multirow{2}{*}{$t$} & 0.05 & 0.25 & 9.84 & 8.82 & 2 & 0.36 & 4.24\\
        & & & 12.27 & 3.26 & 2 & 1.18 & 1.40\\\hline
        \multirow{2}{*}{$T_0$} & 2 & 5 & 9.50 & 8.72 & 2 & 0.36 & 4.19\\
        & & & 13.10 & 2.92 & 2 & 1.07 & 1.22\\\hline
        \multirow{2}{*}{$T_1$} & 2 & 5 & 12.03 & 8.26 & 2 & 0.40 & 3.96\\
        & & & 14.84 & 3.14 & 2 & 1.08 & 1.34\\\hline
        \multirow{2}{*}{$T_2$} & 2 & 5 & 12.03 & 8.26 & 2 & 0.40 & 3.96\\
        & & & 14.84 & 3.14 & 2 & 1.08 & 1.34\\\hline
    \end{tabular}
    \label{table 4}
\end{table}

\section{Benefits and Practical Applications}\label{Section 7}
The multi-objective economic statistical design of the CUSUM chart focuses on optimizing the control chart's parameters by balancing both economic and statistical aspects simultaneously. This allows for effective monitoring and detection of small shifts, often referred to as drifts, in a process. Here are a few potential practical areas of application of this approach:\\
\textbf{1. Manufacturing Process Control:-} The CUSUM chart can very much be used for exercising control in production processes to detect small shifts or drifts in product quality characteristics such as weight, thickness, or other various dimensions. The early detection of process deviations ensures minimal waste, reduced rework costs, and better product quality. The multi-objective economic statistical design of the CUSUM chart helps in balancing the cost of inspections with the cost of defective products.\\
\textbf{2. Chemical and Pharmaceutical Industry:-} The CUSUM chart can also be used for monitoring the concentration of chemicals or ingredients in continuous processes like pharmaceuticals and chemical. The multi-objective economic statistical design of the CUSUM chart helps in maintaining product consistency by detecting small shifts in the chemical mixture or reaction parameters and also helps in reducing the risk of costly recalls.\\
\textbf{3. Healthcare and Medical Processes:-} The CUSUM chart has potential to be used for monitoring vital signs of patients, such as, blood pressure, blood sugar levels, or other health metrics, to detect abnormal trends. The early detection of health related abnormalities reduces the risk of aggravation and the multi-objective economic design of the CUSUM chart can help healthcare providers balance the cost of frequent monitoring with timely interventions ensuring efficient use of the resources.

The application of the multi-objective economic statistical design of the CUSUM chart can be extended to other arenas like Food \& Beverage Industry, Automotive Industry, Supply Chain \& Inventory Management, Financial Market Monitoring, Telecommunications \& Network Monitoring, Environmental Monitoring, and Energy \& Utility Sector.

\section{Conclusions}\label{Section 8}
In this paper, a multi-objective economic statistical design of the CUSUM chart is proposed. The expected cost per cycle ($C_E$) and out-of-control Average Run Length ($ARL_\delta$) are considered as two objectives. This multi-objective problem is then solved with the help of the Non-dominated Sorting Generating Algorithm II (NSGA II). Since there is no research article on the multi-objective design of the CUSUM chart, results are compared with the results of a single objective design proposed by \cite{2011}. The minimum value of $C_E$ obtained by the proposed approach is \$9.50 and the corresponding optimal parameters are $n$ = 2, $h$ = 0.36 hours, and $H$ = 4.19 units. The respective optimal values of $C_E$, $n$, $h$, and $H$ under the normality condition obtained by \cite{2011} are \$16.78, 2, 0.85 hours, and 1.69 units. The proposed multi-objective economic statistical approach reduces the expected cost per cycle ($C_E)$ by 43.39\%. In a practical situation pertaining to an industry, if the cost and time parameters are adequately estimated, the demonstrated algorithm and pseudocode can suitably be used for the effective implementation of the CUSUM control chart.

\bibliographystyle{apalike}
\bibliography{interactapasample}
\section*{Appendix}
\section{Appendix}
\begin{figure}[ht!]
\centering
\begin{minipage}{0.4\textwidth}
    \centering
    \subfigure[Each parameter with low level i.e. for $\delta$ = 1.0.]{\includegraphics[width=7.2cm,height=6cm]{Figure_1.png}}
\end{minipage}
\hfill
\begin{minipage}{0.4\textwidth}
    \centering
    \subfigure[For $\delta$ = 1.5.]{\includegraphics[width=7.2cm,height=6cm]{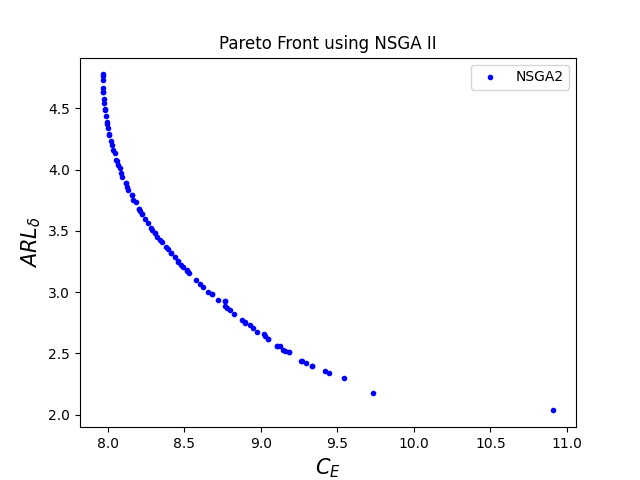}}
\end{minipage}
\hfill
\begin{minipage}{0.4\textwidth}
    \centering
    \subfigure[For $\delta$ = 2.0.]{\includegraphics[width=7.2cm,height=6cm]{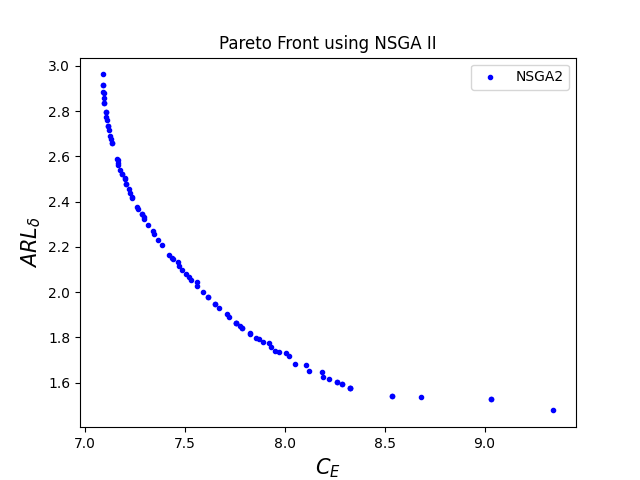}}
\end{minipage}
        \hfill
\begin{minipage}{0.4\textwidth}
    \centering
    \subfigure[For $\delta$ =2.5.]{\includegraphics[width=7.2cm,height=6cm]{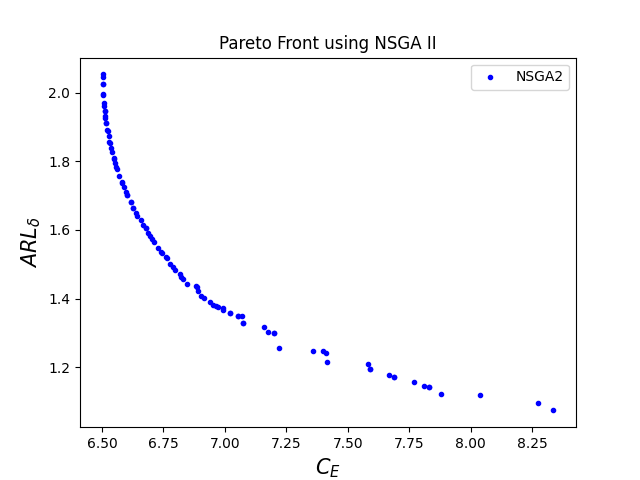}}
\end{minipage}
\caption{Optimal Pareto fronts for different values of $\delta$.}
\label{figure 2}
\end{figure}

The corresponding optimal Pareto fronts for different values of $\delta$ are given in figure \ref{figure 2}.

\begin{figure}
\centering
\begin{minipage}{0.4\textwidth}
    \centering
    \subfigure[$C_0$ with high level.]{\includegraphics[width=6.2cm,height=4.5cm]{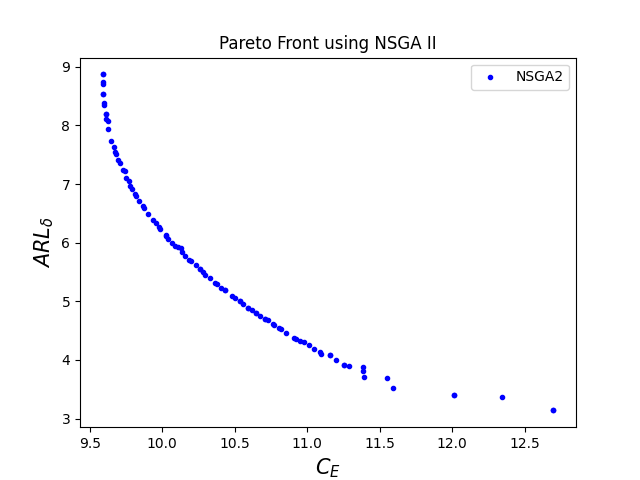}}
\end{minipage}
       \hfill
\begin{minipage}{0.4\textwidth}
    \centering
    \subfigure[$C_1$ with high level.]{\includegraphics[width=6.2cm,height=4.5cm]{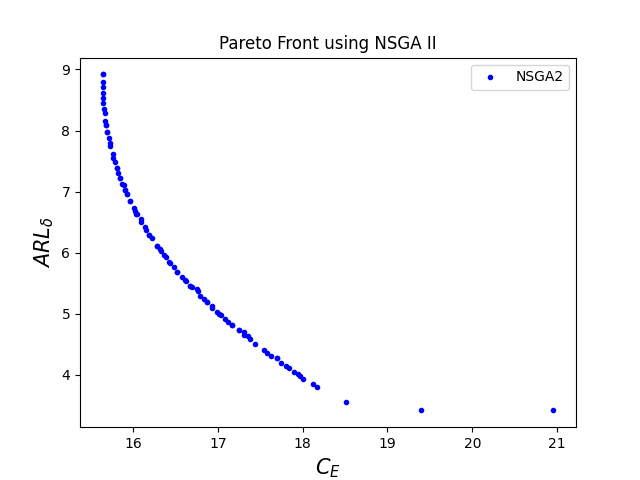}}
\end{minipage}
       \hfill
\begin{minipage}{0.4\textwidth}
    \centering
    \subfigure[$W$ with high level.]{\includegraphics[width=6.2cm,height=4.5cm]{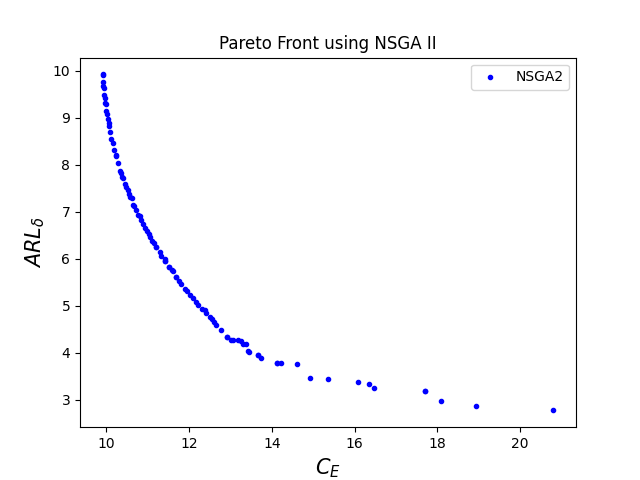}}
\end{minipage}
\hfill
\begin{minipage}{0.4\textwidth}
    \centering
    \subfigure[$Y$ with high level.]{\includegraphics[width=6.2cm,height=4.5cm]{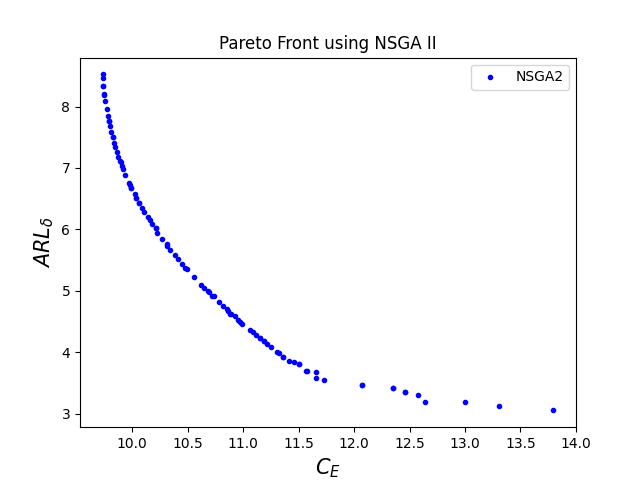}}
\end{minipage}
\hfill
\centering
\begin{minipage}{0.4\textwidth}
    \centering
    \subfigure[$d$ with high level.]{\includegraphics[width=6.2cm,height=4.5cm]{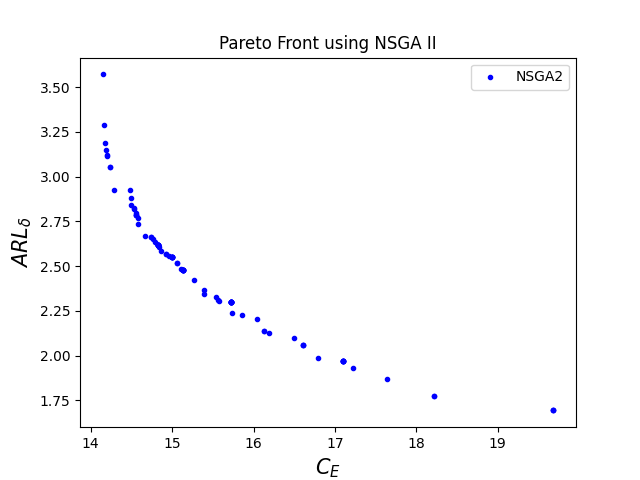}}
\end{minipage}
        \hfill
\begin{minipage}{0.4\textwidth}
    \centering
    \subfigure[$y$ with high level.]{\includegraphics[width=6.2cm,height=4.55cm]{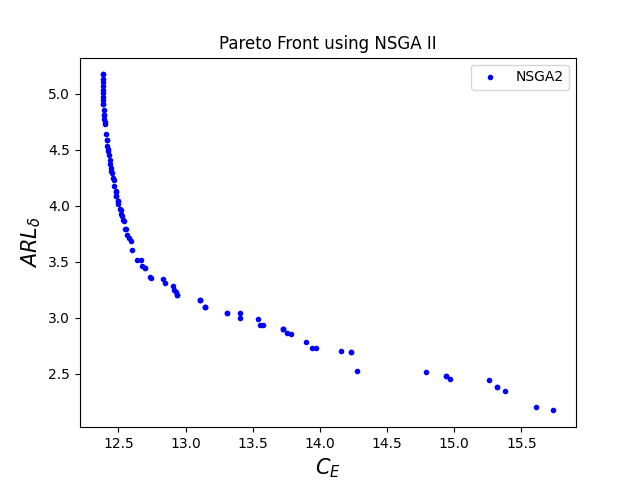}}
\end{minipage}
\caption{Optimal Pareto fronts with high levels of cost parameters.}
\label{figure 3}
\end{figure}

\begin{figure}[ht!]
\centering
\begin{minipage}{0.4\textwidth}
    \centering
    \subfigure[$\lambda$ with high level.]{\includegraphics[width=6.2cm,height=5cm]{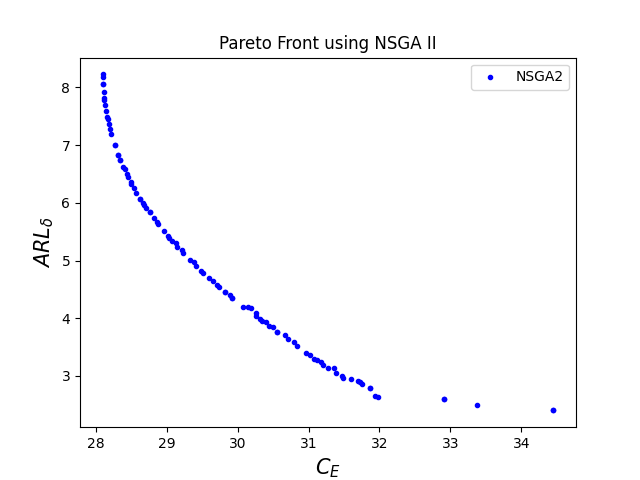}}
\end{minipage}
       \hfill
\begin{minipage}{0.4\textwidth}
    \centering
    \subfigure[$t$ with high level.]{\includegraphics[width=6.2cm,height=5cm]{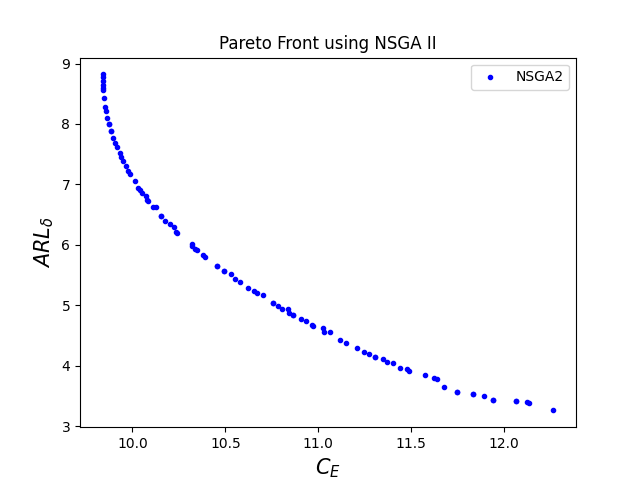}}
\end{minipage}
        \hfill
\begin{minipage}{0.4\textwidth}
    \centering
    \subfigure[$T_0$ with high level.]{\includegraphics[width=6.2cm,height=5cm]{Figure_1.png}}
\end{minipage}
       \hfill
\begin{minipage}{0.4\textwidth}
    \centering
    \subfigure[$T_1$ with high level.]{\includegraphics[width=6.2cm,height=5cm]{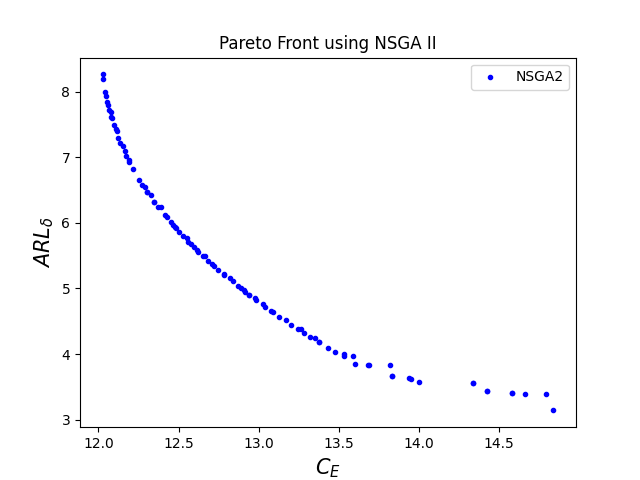}}
\end{minipage}
       \hfill
\begin{minipage}{0.4\textwidth}
    \centering
    \subfigure[$T_2$ with high level.]{\includegraphics[width=6.2cm,height=5cm]{Figure_13.png}}
\end{minipage}
\caption{Optimal Pareto fronts with high levels of time-related input parameters.}
\label{figure 4}
\end{figure}

The corresponding optimal Pareto fronts for high levels of cost parameters are given in figure \ref{figure 3}.\leavevmode\newline\newline\newline\newline\newline\newline
\newpage
The corresponding optimal Pareto fronts for high levels of time-related parameters are given in figure \ref{figure 4}.
\end{document}